\documentclass[12pt]{article}
\usepackage{epsfig}
\if@twoside  
\medskipamount

\def\Journal#1#2#3#4{{#1}{\bf #2}, #4 (#3)}

\def\NPA{{\em Nucl.\ Phys.} A}
\def\PLB{{\em Phys.\ Lett.} B}
\def\PRL{\em Phys.\ Rev.\ Lett. }

\def\PRC{{\em Phys.\ Rev.} C}
\def\PRD{{\em Phys.\ Rev.} D}

\def\ARNPS{\em Annu.\ Rev.\ Nucl.\ Part.\ Sci. }

\newcommand{\be}{\begin{eqnarray}}

\newcommand{\ee}{\end{eqnarray}}
\begin{document}

\thispagestyle{empty}
                           
\begin{flushright}{\sc LBNL-42028}\\\end{flushright}
\vspace{1in}
\begin{center}{\Large \bf { Bremsstrahlung Dileptons in Ultra-relativistic 
Heavy Ion Collisions}}\\
\vspace{1in}
{\large  J. Jalilian-Marian and V. Koch}\\
\vspace{.2in}
{\it  Nuclear Theory Group \\
Nuclear Science Division \\
Lawrence Berkeley National Laboratory\\ 
          Berkeley, CA 94720}\\
\end{center}

\vspace*{25mm}

\begin{abstract}
\baselineskip=18pt

We consider production of dilepton pairs through coherent 
electro-magnetic radiation during nuclear collisions. We show
that the number of pairs produced through bremsstrahlung is about two orders of
magnitude smaller than the yield measured by the CERES
collaboration. Therefore, coherent bremsstrahlung can be ruled out as an
explanation for the observed enhancement of low mass 
dileptons in CERES and HELIOS.  

\end{abstract}

\vspace*{5mm}

%\end{titlepage}
 
\newpage

%\section{Introduction}

Production of low mass dileptons in ultra-relativistic heavy ion collisions
is considered a useful probe of possible in-medium changes of hadrons due to
the onset of chiral restoration (see~\cite{KKL} for a recent review). 
Dilepton are penetrating probes, i.e. once
produced they do not re-interact with the hadronic environment and thus,
provide information about the early stages of the collision where high
temperatures and densities are reached. Since vector mesons such as  
$\rho$, $\omega$  and $\Phi$ have direct decay channels into dileptons,
possible in-medium changes of the vector mesons masses can be observed 
in the dilepton invariant mass spectrum.

There has been a major surge of interest in dilepton mass spectrum
due to a recent experimental observation of an enhanced dilepton 
yield at invariant masses of about $ 400 \rm MeV$ in ${\rm S}$ + ${\rm Au}$ 
and ${\rm Pb}$ + ${\rm Au}$ collisions in SPS compared to proton 
induced collisions 
as reported by CERES collaboration~\cite{CERES}. There is also
a similar enhancement reported by HELIOS collaboration at more
forward rapidities~\cite{HELIOS}. Although the decay of the final state
hadrons explain the data well in proton induced collisions,   
they can not explain the current SPS data for ${\rm S}$ + ${\rm Au}$ 
and ${\rm Pb}$ + ${\rm Au}$
in the mass region around $\sim 400 \, \rm MeV$ .

A major source of this enhancement is simply pion annihilation which is not
present in the proton-nucleus system. Including pion annihilation and
contributions from other hadronic reactions the calculations reach the 
lower end of the sum of statistical and systematic errors of the CERES data 
(see \cite{drees96} for a compilation of different calculations). 
An additional enhancement can be achieved~\cite{LKB,CEK}
if one assumes that the mass of the $\rho$ 
is lowered according to the conjectures of \cite{BR91}. 
Another possible enhancement arises from an in-medium 
modified pion dispersion relation.
While this effect is small if one considers the modification of the 
pion dispersion in a pion gas \cite{KS96}, it is considerably larger if, 
in addition, one takes the effect of baryons into account~\cite{RCW}. 
Consistency of the latter scenario with the observed pion 
to baryon ratio is presently debated.

Another source of dilepton production is simply the bremsstrahlung due to the
deceleration of the incoming nuclei during the collision.
Motivated by a recent result of Mishustin et.al.~\cite{MSSG}, 
we investigate whether this source could account for part of the 
observed enhancement of dilepton numbers as reported by CERES 
collaboration.
In reference~\cite{MSSG} only the production of $\omega$ mesons due to
the deceleration and their subsequent decay into dileptons has been taken 
into account. This corresponds to considering the iso-scalar part of the
electro-magnetic current. However, one also has to take into account the 
iso-vector part, i.e. production of $\rho$-mesons and their subsequent 
decay into dileptons. These two amplitudes, iso-scalar and iso-vector, 
interfere destructively and as a result the dilepton production 
cross section from the coherent deceleration scales like the square 
of the charge and not the square of the baryon number, as assumed 
in~\cite{MSSG}. We should also note that the point like limit of this 
process addressed here has been considered in~\cite{LBGS}. 

As already mentioned, one expects that the coherent  
radiation will be enhanced  by a factor of $Z_{1}Z_{2}$ in nuclear
as compared to proton induced collisions.  
Here $Z_{1}$ and $Z_{2}$ are the atomic numbers of the colliding nuclei. 
The enhancement factor $Z_1Z_2$  follows from assuming coherent 
radiation off of charged nuclei. This assumption
is an approximation which should be valid when one considers photon
virtualities (invariant dilepton masses) which are much smaller than
inverse size of individual nucleons. Therefore, as one goes to higher
and higher invariant dilepton masses, this approximation will cease
to be valid. Also, the ratio of photon virtuality to center of mass
energy of the collision should be small so that there is also 
longitudinal coherence. Even though incoherent radiation
will become as important and eventually dominate over coherent 
radiation as one goes to higher and higher masses, the coherent
radiation will always be there as a background and so therefore, 
it should be understood. Here we try to provide an upper limit on this 
background.

\section{Radiation from Decelerating Nuclei}

Let us consider a typical ultra-relativistic nuclear collision where
nuclei A and B move towards each other with very high but constant velocities.
In order to simplify the extremely complicated process of nuclear
collisions, we will assume that the main effect of the collision
is deceleration of each nucleus while they are passing through
each other. Since electric charges in the nuclei are decelerated, they
emit photons. All time like photons with virtuality $q^2 > 0$ which 
subsequently decay into dileptons are emitted during this time. 
After a passing time $t \sim {{R_{A}+R_{B}} \over {\gamma}}$, 
they move on with reduced but constant velocity. We will ignore 
all subtleties associated with expansion of nuclei in the transverse 
direction during the collision and, for simplicity, 
assume a Gaussian form for the 
charge distribution. 

We can relate the number of dileptons produced in this process to
the Fourier transform of the correlator of the electro-magnetic
currents of the colliding nuclei. It is given by~\cite{MT}

\be
{dN_{l^+ l^-} \over d^{4}p}= { \alpha^2 \over 6\pi^3}
{1 \over p^4} (p^{\mu} p^{\nu} - p^2 g^{\mu \nu}) W_{\mu \nu}(p)
\label{eq:current}
\ee
where $p^{\mu}$ is the virtual photon momentum, $\alpha = 1/137$
is the electro-magnetic coupling constant and $W_{\mu \nu}$ is the
Fourier transform of the product of the electro-magnetic currents:
\be
W_{\mu \nu}(p)=\int d^4x d^4y e^{-ip(x-y)}
J_{\mu}(x) J^{\dagger}_{\nu}(y).
\label{eq:corelator}
\ee

Our task is now simple, we just need to write down an electro-magnetic
current corresponding to an extended charge density with (proper) time
dependent velocity. It is~\cite{ROHR}: 

\be
J_{\mu}^{R}(x)=\int d\tau v_{\mu}(\tau) 
\delta \big[v_{\nu}(x^{\nu} - z^{\nu}(\tau))\big] 
\bigg[1 + a_{\nu}(x^{\nu} - z^{\nu}(\tau))\bigg] 
f[(x-z)^2].
\label{eq:rightcurrent}
\ee
with
\be
z_{0}(\tau)=z_{0}(T) + \int^{\tau}_{T} d\tau^{\prime} 
\gamma(\tau^{\prime}) \nonumber
\ee
and 
\be
z_{3}(\tau)=z_{3}(T) + \int^{\tau}_{T} d\tau^{\prime} 
\gamma(\tau^{\prime})\beta(\tau^{\prime}). \nonumber
\ee
We choose the initial time $T=-\infty$ for convenience. $f[(x-z)^2]$ is
a properly normalized but otherwise arbitrary charge profile at this point.

Here, $\tau$ is the proper time and $a_{\mu}={dv_{\mu} \over d\tau}$ is the
corresponding acceleration (deceleration) with
\be
v_{\mu}(\tau)= \gamma (\tau)
\left( \begin{array}{c}
1 \\
0 \\
0 \\
\beta (\tau)
\end{array}\right)
\label{eq:velocity}
\ee 
being the four velocity 
and $\gamma(\tau)=[1-\beta^2(\tau)]^{-1/2}$. It is easy to verify 
that this current is conserved. The acceleration term $a_{\nu}$ 
looks peculiar and arises out of a consistent application of the
concept of simultaneity in special relativity to an extended (but
still rigid) charged object and is essential for current conservation 
in the case of a spatially extended charge distribution (see~\cite{ROHR}
for a nice illustration of this). 
It will drop out when one takes the point charge limit of this expression.

Expression~(\ref{eq:rightcurrent}) is the current for a charged
nucleus moving from left ($z=-\infty$) to right ($z=+\infty$)
with velocity $\beta$. In order to get the current of a charged
nucleus moving from right to left,
we simply take $\beta \rightarrow -\beta$ in the current of the 
right moving nucleus. The total current to be used 
in~(\ref{eq:corelator}) is the 
sum of the currents of the right moving and left moving nuclei. 
The last step is to determine the velocity $\beta$.
We will assume that both nuclei have a constant initial velocity
$\beta_i$ until they collide at $\tau=\tau_i$. During the collision, from
$\tau_i$ to $\tau_f$, velocity changes in a non-trivial way. After 
$\tau=\tau_f$, both nuclei have again a constant but reduced velocity 
$\beta_f$.

To proceed further, we need to take a specific form for the 
nuclear charge  distribution $f[(x-z)^2]$. For simplicity we  
will use a Gaussian profile

\be
f[(x-z)^2]=\rho_0 
\exp{\bigg[{{(t-z_{0}(\tau))^2 - (\vec{x} - 
\vec{z}(\tau))^2} \over 
2\sigma^2}\bigg]}
\ee
Here $\rho_0$ and $\sigma$ are related to atomic number $Z$ and
radius of nucleus $R$ by
\be
\rho_0 = Z(2\pi R^2/3)^{-{3 \over 2}}, \,\,
\sigma^2 = {1 \over 3} R^2
\nonumber
\ee

A more realistic profile would be a  Woods-Saxon shape. However, since we want
to provide an upper limit of the bremsstrahlung contribution, Gaussian 
distribution is a
reasonable approximation. For a given charge, it is more narrow and will lead
to a larger dilepton yield at finite momentum than the corresponding
Woods-Saxon profile.

In order to model the deceleration phase, one could take the velocity 
during the collision to be a linearly
changing function of time but it is perhaps more realistic
to take the rapidity, rather than the velocity, to be a linearly 
changing function of time. We therefore take the rapidity during the 
collision to be
\be
y(\tau)= y_i + { \Delta y \over \Delta \tau} \tau
\label{eq:rapidity}
\ee
where $\beta (\tau) = \tanh[y(\tau)]$, $\Delta y= y_f - y_i$, 
$\Delta \tau = \tau_f - \tau_i$ 
and the initial and final rapidities and times are to be determined
by experimental considerations.

Fourier transforming the current to momentum space, dividing the
proper time interval into three different regions and performing
the proper time integration in the initial and final stages where
the velocity is constant, we get:
\be
\!\!\!\!J^{R}_{3}(p)\!\!\!\!&=&\!\!\!\!Z\exp
{\bigg[-{1 \over 6} R^2 p_t^2\bigg] } 
\Bigg\{
\beta_i {\exp{\bigg[-{1 \over 6} R^2 \gamma_i^2 (\beta_ip^0 -p^3)^2\bigg]}
\over i[p^0 -\beta_ip_3 -i\epsilon]} \nonumber \\
&-&
\beta_f {\exp{\bigg[-{1 \over 6} R^2 \gamma_f^2 (\beta_fp^0 -p^3)^2 +
i{\Delta\tau \over \Delta y}[p^0(\gamma_f\beta_f - \gamma_i\beta_i) - 
p^3(\gamma_f - \gamma_i)]\bigg]}
\over i[p^0 -\beta_fp_3 +i\epsilon]} \nonumber \\
&+&
\!\!\!\!\int^{\tau_f}_{0} \gamma(\tau)\beta(\tau) 
\bigg[1 - i {R^2 \over 3}{\Delta y \over \Delta \tau}\gamma(\tau)
[\beta(\tau)p^0 - p^3]\bigg] \nonumber \\
&\times&
\!\!\!\!\exp{\bigg[\!\!-\!{1 \over 6} R^2\gamma^2(\tau)
[\beta(\tau)p^0 - p^3]^2 + i 
{\Delta \tau \over \Delta y}[p^0(\gamma(\tau)\beta(\tau) - \gamma_i\beta_i)
- p^3(\gamma(\tau) - \gamma_i)]\bigg]} \Bigg\}
\label{eq:currentft}
\ee 
where we have set $\tau_i =0$ without loss of generality. 
The first two terms (proportional to $\beta_i$ and $\beta_f$) 
correspond to the contribution of the initial
and final stages of the collision respectively while the last
term is the contribution of the time interval when the nuclei are
passing through each other and has to be evaluated numerically. 
Also, in the last term corresponding to the contribution of the deceleration
region, extended structure of the source is responsible for the second term
(proportional to area of the nucleus, $R^2$) in the bracket :
\be
\bigg[1 - i {R^2 \over 3}{\Delta y \over \Delta \tau}\gamma(\tau)
[\beta(\tau)p^0 - p^3]\bigg] 
\label{eq:bracket}
\ee
and will be absent in collision of point charges.

Using current conservation $p_0 J_0(p) = p_3 J_3(p)$ 
and also $J_{\mu} = J_{\mu}^{R} + J_{\mu}^{L}$, we then have 
\be
{d^{4} N_{l^+ l^-} \over dM dy d^2p_t}= { \alpha^2 \over 6\pi^3}
{1 \over M} \bigg[ 1- {p^2_3 \over p^2_0}\bigg] 
J_3(p) J_3^{\dagger}(p).
\ee
As a consistency check, it is easy to show that one does not get
either physical (light like) or time like photons without acceleration. 

Below, we plot our results for the number of dileptons produced
for two different values of collision times with and without the 
CERES acceptance cuts for $\Delta y = 2.4$~\cite{NA}. From here, 
it is clear that bremsstrahlung
is irrelevant for the observed enhancement of dilepton spectrum
even with our Gaussian charge distribution which would clearly over
estimate the number of produced dileptons and that a different 
mechanism is needed. Variation of our parameters like rapidity
change $\Delta y$ and collision time $\Delta \tau$ does not 
change our results appreciably even in the extreme case of 
$\Delta y = y_i$\, i.e. full stoppage of nuclei after collision.    

We also plot the transverse momentum spectrum
of the produced dileptons, integrated over rapidity and dilepton 
invariant mass range $[200-600 \,\rm MeV]$ and compare it with CERES 
data. Both graphs
have the peculiarity that as one increases the time it takes 
the nuclei to pass through each other from $1 \,\rm fm$ to $4 \,\rm fm$, 
the number of dileptons increases in contrast to what one
expects intuitively (at least for high invariant masses).
This is again due to the extended structure of the source. 
Unlike the case of a point charge, we have an inherent 
scale in the problem, namely $R$, the nuclear radius. 

\begin{figure}[htp]
\centering
\setlength{\epsfxsize=10cm}
\centerline{\epsffile{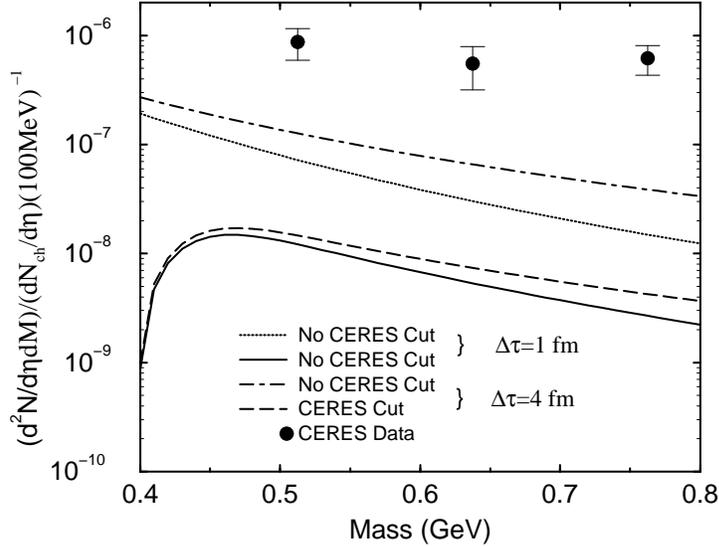}}
\caption{ Number of dileptons produced for $1\, \rm fm$ and $4\, \rm fm$ 
collision times with and without CERES cuts.} 
\label{fig:mspect}
\end{figure}

Physically, there are two competing effects which are responsible 
for this initial increase in the number of emitted dileptons
as one increases the collision time. The first effect is just what 
one expects; as charges decelerate over a longer time 
interval, they emit fewer dileptons. The second effect is due to
the finite time required to build up coherence among different
parts of the extended source. In other words, if the nucleus
decelerates too quickly, there is no time for the different points
in the extended source to communicate and coherent emission takes place 
only from a fracture of the source which has had time to react. As we increase
the collision (deceleration) time, we increase the fraction of the nucleus
which can be considered a coherent source  and as a 
result, we have more dileptons emitted. After some time,
the whole nucleus is emitting dileptons coherently after which the 
first effect takes over and the number of dileptons starts decreasing 
as we increase the collision time any further. 

%Formaly,
%this is due to the fact that expression~(\ref{eq:currentft}),
%considered as a function of $\tau_f$, develops a maximum at some
%$\tau_f \sim R$ due to presence of the second term 
%in~(\ref{eq:bracket}) while in the case of point charges, the expression
%for the number of dilepton pairs is a monotonously decreasing function
%of $\tau_f$. The number of dileptons increases as we increase
%$\tau_f$ from zero up to this point. If we increase the time any further, 
%dilepton numbers will decrease and eventually go to zero as 
%$\tau_f \rightarrow \infty$ in agreement with point charge results 
%and intuitive expectations.  

\vspace{.5in}
\begin{figure}[htp]
\centering
\setlength{\epsfxsize=10cm}
\centerline{\epsffile{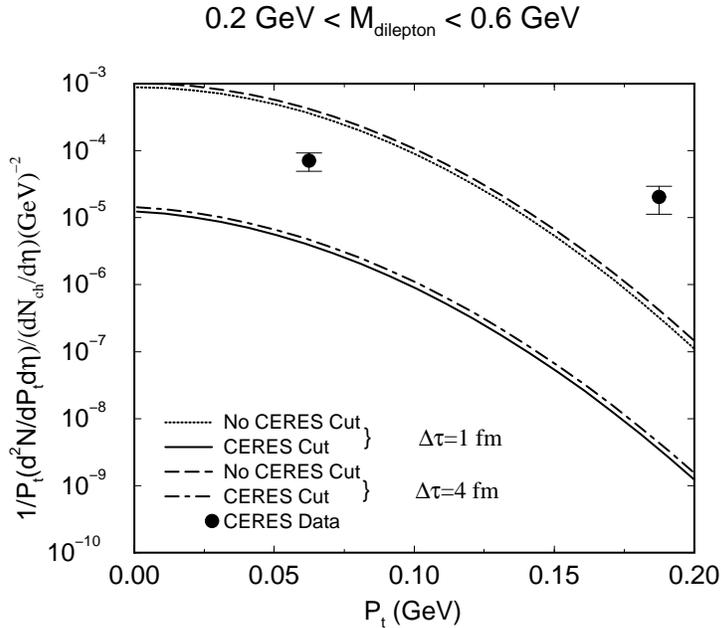}}
\caption{ $P_t$ spectrum of dileptons produced for $1 \,\rm fm$ and 
$4\, \rm fm$ with and without CERES cuts integrated over rapidity and 
mass in the mass range between $200$ and $600$ MeV.} 
\label{fig:nvspt}

\end{figure}

It is interesting to note that dilepton mass spectrum due to 
deceleration of a point 
charge (not shown here) shows a (modulated) periodicity which
depends on the time it takes the nuclei to pass
through each other and in principle could be used to
determine this time. However, this structure is totally wiped out
by the Gaussian charge distribution and it is unlikely that it
could be experimentally useful in determining the collision time for
a realistic charge distribution.

In summary, we consider production of dilepton pairs due to coherent 
bremsstrahlung in ultra-relativistic heavy ion collisions. We 
provide an estimate of the upper limit for this contribution using
a Gaussian charge distribution. We find coherent bremsstrahlung 
to be a negligible source for dileptons.

This work was supported by the Director, Office of Energy Research, 
Office of High Energy and Nuclear Physics Division of the Department 
of Energy, under contract No. DE-AC03-76SF00098 and DE-FG02-87ER40328.

\vspace{.3in}
\leftline{\bf References}

\renewenvironment{thebibliography}[1]
        {\begin{list}{[$\,$\arabic{enumi}$\,$]}  % {\arabic{enumi}.}
        {\usecounter{enumi}\setlength{\parsep}{0pt}
         \setlength{\itemsep}{0pt}  \renewcommand{\baselinestretch}{1.2}
         \settowidth
        {\labelwidth}{#1 ~ ~}\sloppy}}{\end{list}}

\end{document}